\documentclass[12pt]{article}
\usepackage{axodraw,bbold}

\catcode`@=12
\begin{document}
\thispagestyle{empty}
\begin{flushright}
OSU-HEP-07-03\\
UCRHEP-T438\\
August 2007\
\end{flushright}
\vspace{0.5in}
\begin{center}
{\LARGE \bf Singlet fermion dark matter and electroweak\\
baryogenesis with radiative neutrino mass\\}
\vspace{0.7in}
{\bf K. S. Babu\\}
\vspace{0.1in}
{\sl Department of Physics, Oklahoma State University,\\
Stillwater, Oklahoma 74078, USA\\}
\vspace{0.2in}
{\bf Ernest Ma\\}
\vspace{0.1in}
{\sl Department of Physics and Astronomy, University of
California,\\ Riverside, California 92521, USA\\}
\vspace{0.7in}
\end{center}

\begin{abstract}\
The model of radiative neutrino mass with dark matter proposed by one of us
is extended to include a real singlet scalar field.  There are then two
important new consequences.  One is the realistic possibility of having the
lightest neutral singlet fermion (instead of the lightest neutral component
of the dark scalar doublet) as the dark matter of the Universe.  The other
is a modification of the effective Higgs potential of the Standard Model,
consistent with electroweak baryogenesis.
\end{abstract}

\newpage
\baselineskip 24pt

With the addition of a second scalar doublet $\eta = (\eta^+,\eta^0)$
[$\eta^0 = (\eta^0_R + i \eta^0_I)/\sqrt{2}$] to
the Standard Model (SM) of quark and lepton interactions, a cornucopia of
new opportunities opens up for the understanding of physics phenomena beyond
the SM.  One possibility \cite{dm78} is that $\eta$ is odd with respect to an
exactly conserved discrete $Z_2$ symmetry, allowing \cite{m06-1,bhr06,m06-2,
m06-3,lnot07,hkmr07,glbe07,ms07,ht07} $\eta^0_R$ or $\eta^0_I$
to be a candidate for the dark matter \cite{bhs05} of the Universe.

If three heavy neutral singlet Majorana fermions $N_{1,2,3}$ are added as
well \cite{m06-1,m06-2,m06-3,ms07}, which are odd under the aforementioned
$Z_2$, then neutrinos acquire radiative seesaw masses through the Yukawa
interactions
\begin{equation}
h_{ij} (\nu_i \eta^0 - l_i \eta^+) N_j + H.c.
\end{equation}
and the mass splitting of $\eta^0_R$ and $\eta^0_I$
from the quartic scalar term
\begin{equation}
{1 \over 2} \lambda_5 (\Phi^\dagger \eta)^2 + H.c.,
\end{equation}
where $\Phi = (\phi^+,\phi^0)$ is the usual SM Higgs doublet with $\langle
\phi^0 \rangle = v/\sqrt{2}$, whereas $\langle \eta^0 \rangle = 0$.  Thus the
neutrinos of this model have no Dirac masses linking them to $N_i$, but
they obtain Majorana masses in one loop given by \cite{m06-1}
\begin{equation}
({\cal M}_\nu)_{ij} = \sum_k {h_{ik} h_{jk} M_k \over 16 \pi^2} \left[
{m_R^2 \over m_R^2-M_k^2} \ln {m_R^2 \over M_k^2} - {m_I^2 \over m_I^2 -
M_k^2} \ln {m_I^2 \over M_k^2} \right].
\end{equation}

Instead of $\eta^0_R$ or $\eta^0_I$, the lightest among
the $N_i$ fermions may now be considered \cite{m06-1,knt03,kms06} as a
dark-matter candidate.  However, if Eq.~(1) is the only interaction of $N_i$,
the requirement of a realistic dark-matter relic abundance is generally in
conflict \cite{kms06} with flavor changing radiative decays such as $\mu
\to e \gamma$, which cannot be alleviated without some degree of fine tuning.

One way to evade the above constraint is to endow $N$ with some other 
interaction, such as an extra gauge $U(1)'$ \cite{ks06}.  Here we propose 
instead the minimal addition of a real singlet scalar
$\chi$ to allow another channel for $NN$ annihilation, thus freeing the
constraint of relic abundance from Eq.~(1).  With smaller values of $h_{ij}$,
flavor changing radiative decays such as $\mu \to e \gamma$ are suitably
suppressed.  At the same time, with the addition of $\chi$, the effective
Higgs potential involving the SM Higgs doublet $\Phi$ is now such
\cite{z93,volkas,gsw05}
that electroweak baryogenesis \cite{krs85} may also become possible, as 
elaborated below.

There are two possible scenarios in which successful baryogenesis could be 
realized within the present model. In the first scenario the singlet scalar 
$\chi$ remains light down to the scale of electroweak symmetry breaking.  
Its interactions with the SM Higgs doublet modifies the condition for 
strongly first-order phase transition and leads to successful baryogenesis 
without contradicting the lower limit on the SM Higgs boson mass \cite{volkas}.
We have nothing new to add for this scenario.  There is a second
possibility where the singlet scalar $\chi$ has a mass larger than the 
electroweak scale.  One can then integrate out $\chi$ from the effective 
low energy theory. The potential for the SM Higgs doublet gets modified 
in this process, enabling successful baryogenesis \cite{gsw05}.

Consider first the most general renormalizable Higgs potential of two
doublets $\Phi = (\phi^+,\phi^0)$, $\eta = (\eta^+,\eta^0)$, and a singlet
$\chi$, where $\eta$ is odd under an extra $Z_2$ as proposed in
Ref.~\cite{m06-1}:
\begin{eqnarray}
V &=& m_1^2 \Phi^\dagger \Phi + m_2^2  \eta^\dagger \eta + {1 \over 2}
m_3^2 \chi^2 + {1 \over 2} \lambda_1 (\Phi^\dagger \Phi)^2
+ {1 \over 2} \lambda_2 (\eta^\dagger \eta)^2 + \lambda_3
(\Phi^\dagger \Phi)(\eta^\dagger \eta) \nonumber \\ &+&
\lambda_4 (\eta^\dagger \Phi)(\Phi^\dagger \eta) + {1 \over 2} \lambda_5
[(\eta^\dagger \Phi)^2 + (\Phi^\dagger \eta)^2] + \mu_1 \chi
(\Phi^\dagger \Phi) + \mu_2 \chi (\eta^\dagger \eta) \nonumber
\\  &+& {1 \over 2} \mu_3 \chi^3 + {1 \over 8} \lambda_6 \chi^4 +
{1 \over 2} \lambda_7 \chi^2 (\Phi^\dagger \Phi) + {1 \over 2} \lambda_8
\chi^2 (\eta^\dagger \eta),
\end{eqnarray}
where $\lambda_5$ has been chosen real without any loss of generality.
To obtain the tree-level effective potential containing only $\Phi$ and
$\eta$, we eliminate $\chi$ by its own equation of motion in powers of
$|\Phi|^2$ and $|\eta|^2$.  We assume that $\langle \chi \rangle = 0$
at this point.  To eighth order in the fields, we then obtain
\begin{eqnarray}
V_{eff} &=& m_1^2 \Phi^\dagger \Phi + m_2^2 \eta^\dagger \eta + {1 \over 2}
\left( \lambda_1 - {\mu_1^2 \over m_3^2} \right) (\Phi^\dagger \Phi)^2
+ {1 \over 2} \left( \lambda_2 - {\mu_2^2 \over m_3^2} \right) (\eta^\dagger
\eta)^2 \nonumber \\ &+& \left( \lambda_3 - {\mu_1 \mu_2 \over m_3^2} \right)
(\Phi^\dagger \Phi)(\eta^\dagger \eta) + \lambda_4 (\eta^\dagger \Phi)
(\Phi^\dagger \eta) + {1 \over 2} \lambda_5
[(\eta^\dagger \Phi)^2 + (\Phi^\dagger \eta)^2] \\
&+& { (\mu_1 \Phi^\dagger \Phi + \mu_2 \eta^\dagger \eta)^2 \over 2 m_3^4}
\left[ \left(
\lambda_7 - {\mu_1 \mu_3 \over m_3^2} \right) \Phi^\dagger \Phi + \left(
\lambda_8 - {\mu_2 \mu_3 \over m_3^2} \right) \eta^\dagger \eta \right]
\nonumber \\
&+& {(\mu_1 \Phi^\dagger \Phi + \mu_2 \eta^\dagger \eta)^2 \over 2 m_3^6}
\left\{ {\lambda_6 (\mu_1 \Phi^\dagger \Phi + \mu_2 \eta^\dagger \eta)^2
\over 4 m_3^2} - \left[ \left( \lambda_7 - {3 \mu_1 \mu_3 \over
2 m_3^2} \right) \Phi^\dagger \Phi + \left( \lambda_8 - {3 \mu_2 \mu_3 \over
2 m_3^2} \right) \eta^\dagger \eta \right]^2 \right\}. \nonumber
\end{eqnarray}
We seek a solution where $Z_2$ is not broken, i.e. $\langle \eta^0 \rangle
= 0$, in which case the effective Higgs potential for $\Phi$ alone is
given by
\begin{eqnarray}
V_{eff} (\eta = 0) &=& m_1^2 |\Phi|^2 + {1 \over 2} \left( \lambda_1 -
{\mu_1^2 \over m_3^2}
\right) |\Phi|^4 + {\mu_1^2 \over 2 m_3^4} \left( \lambda_7 - {\mu_1 \mu_3
\over m_3^2} \right) |\Phi|^6 \nonumber \\
&+& {\mu_1^2 \over 2 m_3^6} \left[ {\lambda_6 \mu_1^2 \over 4 m_3^2} -
\left( \lambda_7 - {3 \mu_1 \mu_3 \over 2 m_3^2} \right)^2 \right] |\Phi|^8.
\end{eqnarray}
This is of the form obtained by Ref.~\cite{gsw05} where the coefficient
of the $|\Phi|^4$ term may be chosen negative to allow for a strong
first-order phase transition required by electroweak baryogenesis.
The numerical conditions have been analyzed fully in Ref.~\cite{gsw05}
and we have nothing to add here.

As for CP violation needed for electroweak baryogenesis, the SM contribution
from the CKM phase is known to be too small.  The new Yukawa couplings 
$h_{ij}$ of
leptons to the $\eta$ doublet (see Eq. (1)) contain new CP violating phases.
However, since $\eta$ does not directly participate in electroweak symmetry
breaking, these phases are unlikely to be significant for baryogenesis.
There is however another source of CP violation in the model -- the strong
CP violation parameter $\overline{\theta}$ \cite{goran}.  This parameter
can be of order unity at
temperatures of order 100 GeV.  The Peccei-Quinn \cite{pq77} mechanism
which solves the strong CP problem indeed assumes the initial value of 
$\overline{\theta}$ to be of order unity.  Once the QCD phase transition 
is turned on, at temperatures of order 1 GeV, the PQ mechanism ensures that 
$\overline{\theta}$ is relaxed dynamically to zero.  The effect of 
$\overline{\theta}$ on electroweak baryogenesis has been
studied in Ref. \cite{kst}, where it is shown that this might be sufficient
for baryogenesis, but in this case the axion cannot be the dark matter.
In our model of course, the lightest $N_i$ remains our choice for dark matter.
(It is also possible to introduce higher dimensional operators 
which violate CP, as in Ref.~\cite{huber}.)

Going back to Eq.~(4), suppose $\chi$ is the remnant of a spontaneously 
broken $U(1)$ symmetry, then it is easy to show that the following 
parameters are related:
\begin{equation}
m_3^2 = \lambda_6 v_S^2, ~~~ \mu_3 = \lambda_6 v_S, ~~~ \mu_1 = \lambda_7 v_S,
~~~ \mu_2 = \lambda_8 v_S.
\end{equation}
In that case, Eq.~(5) reduces exactly to
\begin{eqnarray}
V_{eff} [U(1)] &=& m_1^2 \Phi^\dagger \Phi + m_2^2 \eta^\dagger \eta +
{1 \over 2} \left( \lambda_1 - {\lambda_7^2 \over \lambda_6} \right)
(\Phi^\dagger \Phi)^2 + {1 \over 2} \left( \lambda_2 - {\lambda_8^2 \over
\lambda_6} \right) (\eta^\dagger \eta)^2 \nonumber \\ &+& \left( \lambda_3 -
{\lambda_7 \lambda_8 \over \lambda_6} \right) (\Phi^\dagger \Phi)
(\eta^\dagger \eta) + \lambda_4 (\eta^\dagger \Phi)
(\Phi^\dagger \eta) + {1 \over 2} \lambda_5
[(\eta^\dagger \Phi)^2 + (\Phi^\dagger \eta)^2],
\end{eqnarray}
i.e. to all orders in $\Phi$ and $\eta$.  The proof is very simple.
It merely comes from the fact that the combination $(v_S + \chi)^2$ always
appears together in $V$.  This also means that the singlet Majoron model
of spontaneous lepton number violation \cite{cmp81} will not generate
a nonzero $|\Phi|^6$ term (or any other beyond $|\Phi|^4$) in Eq.~(6).

However, if the $U(1)$ lepton symmetry is spontaneously broken at the TeV 
scale, with $S = (\chi + v_S)\exp(iJ/v_S)/\sqrt{2}$ but with 
$m_\chi$ at the electroweak scale, 
then electroweak baryogenesis can be successful \cite{volkas}. Even though 
our model has a second Higgs doublet $\eta$, TeV scale lepton number 
violation is consistent with all experimental and astrophysical data.  This 
is because $\left\langle \eta^0 \right\rangle = 0$  and the Majoron $J$ 
resides entirely in the (complex) singlet $S$ \cite{cmp81}.
However, our model is different from the singlet
Majoron model in one important respect.  The charged  scalar $\eta^\pm$ induces
charged-lepton couplings to the Majoron in our model, which is stronger
than those in the singlet Majoron model. We find these couplings to be
\begin{equation}
{\cal L}^{ij}_{\rm eff} = i \sum_k{h_{ik}^* h_{jk} \over 32 \sqrt{2} 
\pi^2 v_S} \left[ {r_k \over 1-r_k}
+ {r_k ~{\rm ln} r_k \over (1-r_k)^2} \right] \{(m_i - m_j) 
\overline{e}_j e_i J + (m_i + m_j) \overline{e}_j
\gamma_5 e_iJ \}
\end{equation}
where $r_k \equiv (M_k^2/M_{\chi^\pm}^2)$ with $M_k$ being the mass of 
$N_k$, and $m_i$ stands for the
mass of the $i$th charged lepton.  Note that these couplings can lead to 
decays such as
$\mu \rightarrow e J$.  The relevant Yukawa coupling can be written, 
for $M_k \sim M_{\chi^\pm}
\sim v$, as $(m_\nu m_\mu/\lambda_5 M_k^2)$ which is of order 
$10^{-15}/\lambda_5$.
(Note that the light neutrino mass is proportional to $\lambda_5$, while 
these induced couplings are not.)
The branching ratio for the decay $\mu \rightarrow e J$ is of order 
$10^{-12}/\lambda_5^2$, which is below the current experimental limit 
for $\lambda_5 \sim 10^{-2} - 1$.  Since the branching ratio is close 
to the current experimental limit, there is some hope that this decay 
may be accessible to the next round of experiments.

Going back to the case of a heavy real singlet $\chi$, 
consider the electroweak symmetry breaking due to Eq.~(6) up to order
$|\Phi|^6$.  Let $\phi^0 = (v + h)/\sqrt{2}$, then \cite{z93}
\begin{equation}
m_h^2 = -4 m_1^2 - \left( \lambda_1 - {\mu_1^2 \over m_3^2} \right) v^2,
\end{equation}
and the cubic interaction
\begin{equation}
\left[ {5 m_h^2 \over 6 v} - {1 \over 3} \left( \lambda_1 - {\mu_1^2
\over m_3^2} \right) v \right] h^3
\end{equation}
appears. The SM is recovered if $m_1^2 = -(\lambda_1 - \mu_1^2/m_3^2)v^2/2$.
Since $\langle \phi^0 \rangle \neq 0$, Eq.~(4) implies that
\begin{equation}
\langle \chi \rangle = {-\mu_1 v^2 \over m_3^2} + {\mu_1 \over m_3^4}
\left( \lambda_7 - {3 \mu_1 \mu_3 \over 2 m_3^2} \right) v^4,
\end{equation}
and $h$ mixes with $\chi$.

Consider now the Yukawa couplings $f_{ij} \chi N_i N_j$.  Unlike the
$h_{ij}$ of Eq.~(1), these are not constrained by flavor changing radiative
decays such as $\mu \to e \gamma$.  The process
\begin{equation}
N_i~N_j \to \chi \to h~h,
\end{equation}
will contribute to the relic abundance of the lightest $N_i$ as dark matter.
Furthermore, the term $\chi (\Phi^\dagger \Phi)$ in $V$ will mix $h$ with
$\chi$.  As a result, the processes
\begin{equation}
N_i~N_j \to h \to W^+~W^-,~Z~Z,~h~h,~{\rm etc.}
\end{equation}
are also possible.  Presumably, the direct detection of the lightest $N_i$
will be from the elastic scattering of $N_i$ off nuclei through $h$
exchange.

Let $N_1$ be the lightest singlet fermion, then its nonrelativistic 
annihilation cross section in the early Universe from Eq.~(13) multiplied 
by its relative velocity is given by
\begin{equation}
\sigma v_{rel} = {f_{11}^2 \over 64 \pi} {\mu_1^2 \sqrt{1-(m_h^2/M_1^2)}
\over (4M_1^2-m_\chi^2)^2}.
\end{equation}
Assuming this to be the dominant contribution to the dark-matter relic
density of the Universe, we need $\sigma v_{rel}$ of order 1 pb \cite{kt90},
which may be obtained for example with $M_1 = 200$ GeV, $m_h = 125$ GeV,
$m_\chi (= m_3) = 500$ GeV, $\mu_1 = 400$ GeV, and $f_{11} = 0.18$.
The heavier $N_{2}$ will decay through Eq.~(1) to $N_1 l_i \bar l_j$ and
$N_3$ to $N_{1,2} \nu_i \bar \nu_j$.

If there is a lepton family symmetry such as $A_4$ \cite{mr01,bmv03} which
makes $h_{ij} = h \delta_{ij}$ in Eq.~(1), then the neutrino mass matrix
of Eq.~(3) is diagonalized by the same unitary transformation $U$ which
diagonalizes the $3 \times 3$ mass matrix ${\cal M}_N$.  In that case,
we have the prediction that the neutrino mass eigenvalues responsible for
neutrino oscillations are given by
\begin{equation}
m_i = {h^2 M_i \over 16 \pi^2} \left[ {m_R^2 \over m_R^2-M_i^2}
\ln {m_R^2 \over M_i^2} - {m_I^2 \over m_I^2 - M_i^2} \ln {m_I^2 \over M_i^2}
\right].
\end{equation}
This may be verifiable experimentally from $\eta$ decay.

With $N_1$ as dark matter, its direct detection becomes very
difficult, because its elastic scattering cross section with nuclei
is only through $h$ exchange \cite{bhr06,lnot07}, with the small effective
coupling $f_{11} \mu_1 v/m_\chi^2 \simeq 0.07$.  Our model has many other
testable predictions, including the production of neutral and charged
scalar particles $\chi,~\eta^0$ and $\eta^\pm$ at the Large Hadron
Collider \cite{cmr07}.

This work was supported in part by the U.~S.~Department of Energy under Grant
Nos.~DE-FG03-94ER40837 and ~DE-FG03-98ER41076.

\newpage

\baselineskip 18pt

\bibliographystyle{unsrt}

\end{document}